\documentclass[11pt]{article}

\newcommand{\be}{\begin{equation}}
\newcommand{\ee}{\end{equation}}
\newcommand{\bc}{\begin{center}}
\newcommand{\ec}{\end{center}}

\usepackage{mathrsfs}
\usepackage{amsfonts}
\usepackage{graphicx}

\begin{document}
\def\theequation{\arabic{section}.\arabic{equation}}
\begin{titlepage}
\title{Embedding black holes and other inhomogeneities in the universe in 
various theories of gravity: a short review}
\author{Valerio Faraoni \\ \\
{\small Department of Physics and Astronomy, Bishop's University} \\
{\small 2600 College Street, Sherbrooke Qu\'ebec, Canada J1M~1Z7}\\
}
\date{}
\maketitle   \thispagestyle{empty}  \vspace*{1truecm}
\begin{abstract} 

The classic black hole mechanics and thermodynamics are formulated for 
stationary black holes with event horizons. Alternative theories of 
gravity of interest for cosmology contain a built-in time-dependent 
cosmological ``constant'' and black holes are not stationary. Realistic 
black holes are anyway dynamical because they interact with astrophysical 
environments or, at a more fundamental level, because of backreaction by 
Hawking radiation. In these situations the teleological concept of event 
horizon fails and apparent or trapping horizons are used instead. Even as 
toy models, black holes embedded in cosmological ``backgrounds'' and other 
inhomogeneous universes constitute an interesting class of solutions of 
various theories of gravity. We discuss the known phenomenology of 
apparent and trapping horizons in these geometries, focusing on 
spherically symmetric inhomogeneous universes.

\end{abstract}
\end{titlepage}   
\clearpage

\section{Introduction}
\label{sec:1}

Black holes are a fundamental prediction of General Relativity (GR) and 
have been the subject of a celebrated theory of dynamics and 
thermodynamics developed in the 1970s \cite{Wald}. The black holes 
considered 
in these studies are stationary solutions of GR. Real black holes, 
however, are non-stationary due to several possible processes:

\begin{itemize}

\item A cosmological background: real black holes are embedded in the 
universe and, although this feature may be irrelevant for their 
astrophysics, their asymptotics are quite relevant in problems of 
principle and in mathematical physics. Alternative theories of gravity 
advocated to explain the present acceleration of the universe without an 
{\em ad hoc} dark energy contain a built-in, time-dependent effective 
cosmological ``constant'' and black holes in these theories are 
asymptotically 
Friedmann-Lema\^itre-Robertson-Walker (FLRW).

\item An astrophysical environment, which could include a companion in a 
binary system (this is the situation which led to the detection 
of gravitational waves in a black hole merger by {\em 
LIGO} \cite{LIGO, LIGO2, LIGO4}), an accretion disk, or spherical 
accretion.

\item Hawking radiation and evaporation affect all 
black holes and are, therefore, important and unavoidable in all problems 
of principle. They become important for hypothetical small black holes in 
their final 
stages or perhaps even for primordial black holes in the early universe.

\end{itemize}

\subsection{A problem of principle}

The non-stationarity of black holes constitutes a problem that may be 
safely ignored in most astrophysical contexts but it becomes 
important, or even crucial, in a fundamental physics context. Let us 
recall a few motivations for the study of dynamical black holes (a more 
extensive discussion can be found in \cite{Sotiriou, mybook}).

The first problem to note is the fact that black holes are embedded in a 
universe, that is, they are not asymptotically flat. Asymptotic flatness 
is an essential ingredient in the proof of the no-hair theorems of 
scalar-tensor gravity, which provide a fundamental understanding of black 
holes in these theories. While there is literature on 
asymptotically 
de Sitter or anti-de Sitter black holes, it is much more realistic to 
consider asymptotically FLRW black 
holes. In the astrophysical context, the cosmological asymptotics 
cannot be neglected for primordial black holes.

Theories of gravity alternative to GR, aiming at explaining the cosmic 
acceleration without dark energy ({\em e.g.}, $f(R)$ gravity \cite{CCT, 
CDTT}) contain a built-in, time-dependent cosmological ``constant''. 
Therefore, black holes in theories of this class which are relevant for 
present-day cosmology are not asymptotically flat. Understanding black 
holes is an important part of understanding a theory of gravity, 
and the FLRW asymptotics are essential here.

Dynamical inhomogeneous spherically symmetric universes are theoretical 
testbeds which allow one to probe  
ideas about the backreaction of inhomogeneities, or the idea that we live 
in a giant void, which have been proposed as alternatives to dark energy 
to explain the cosmic acceleration (see the review 
\cite{BolejkoCelerier}).

Black hole mechanics and thermodynamics were developed for stationary 
black holes with event horizons, which are null surfaces. Realistic black 
holes are instead dynamical and have apparent horizons, which are timelike 
or spacelike surfaces and they may change their causal nature during 
the evolution.

If primordial black holes were formed in the early universe, at some time 
they could have had a size non-negligible in comparison with the Hubble 
scale $\sim H^{-1}$ and very dynamical horizons. A relevant question is 
how fast they could have accreted material and have grown, even though 
primordial black holes are unlikely to be a dominant component of dark 
matter today. Another interesting problem of principle is the accretion of 
dark energy, especially phantom energy, onto black holes 
\cite{Babichevetal04, ChenJing05, IzquierdoPavon06, PachecoHorvath07, 
MaedaHaradaCarr08, GaoChenVFShen08, Guarientoetal08, Sun08, Sun09, 
GonzalezGuzman09, Heetal09, Babichevetal11, Nouicer11, ChadburnGregory13}. 
The simplest model is, of course, spherical accretion but this is already 
a very non-trivial problem requiring the introduction of some toy model.

Another motivation of principle is the study of the spatial variation of 
the fundamental constants of physics. This idea goes back to 
Dirac and is partially implemented in scalar-tensor gravity, in which the 
gravitational coupling strength $G_{eff}$ becomes a dynamical field 
(roughly speaking, the inverse of the Brans-Dicke scalar field $\phi$). 
While the time variation of $G_{eff}$ has been studied extensively in 
scalar-tensor cosmology, its spatial variation requires the analysis of 
inhomogeneous universes, which have been the subject of a much smaller 
literature.

\subsection{A practical problem}

There is a much more practical and urgent problem in gravitational 
physics: highly dynamical black holes are very important in certain 
astrophysical 
applications. The first direct detection of gravitational waves by the 
{\em LIGO} interferometers involved a black hole merger \cite{LIGO}. In 
the final approach and during coalescence of these black holes, the 
spacetime is extremely 
dynamical. The signal to noise ratio in {\em LIGO} interferometers is 
typically low 
and one needs to match the signal to templates calculated theoretically. 
Numerical simulations of black hole mergers are used to generate banks of 
templates of gravitational waveforms for {\em LIGO} detection. These 
simulations do not rely on event horizons, but they use instead apparent 
and trapping horizons. Event horizons are essentially useless for 
practical purposes such as the generation of these banks of templates. 
Even though an unusually loud gravitational wave signal from a binary 
black hole merger can be spotted without templates, the determination of 
the orbital parameters of the black hole binary requires fitting the 
signal with templates computed by using apparent horizons. In the 
calculations producing these templates, ``black holes'' 
are identified with outermost marginally trapped surfaces and apparent 
horizons ({\em e.g.}, \cite{Thornburg07, BaumgarteShapiro03, 
ChuPfeifferCohen11, CookWangSperhake}). It is fair to say that, in 
practice, in astrophysics 
we use apparent horizons and not event horizons.

It is impossible to discuss dynamical black holes without speaking of 
apparent horizons. Therefore, in the next section we introduce apparent 
and 
trapping horizons and their problems. Sec.~\ref{sec:3} presents a 
selection of exact solutions of various theories of gravity describing 
black holes embedded in cosmological ``backgrounds''.\footnote{We use the 
word ``background'' in quotation marks because the non-linearity of the 
field equations forbids splitting the spacetime metric into a background 
and deviations from it in a covariant way (except for algebraically 
special geometries such as Kerr-Schild metrics).} The last section  
outlines open problems.

\section{Apparent horizons and their problems}
\label{sec:2}

In the words of Rindler \cite{Rindler56}, a horizon is {\em a frontier 
between things observable and things unobservable}.  The horizon, a 
product of strong gravity, characterizes a black hole.

There are several kinds of ``textbook'' horizons: Rindler horizons 
associated with uniform acceleration even in flat space, black hole 
horizons, and 
cosmological horizons. There are event, Killing, inner, outer, Cauchy, 
apparent, trapping, quasi-local, isolated, dynamical, and slowly evolving 
horizons \cite{Poisson, Wald, Booth, Nielsen, AshtekarKrishnan, 
GourghoulhonJaramillo08}. Various horizon notions coincide for stationary 
black holes.

In black hole thermodynamics, if the ``background'' is not Minkowskian, 
the internal energy appearing in the first law of thermodynamics must be 
defined carefully. This is identified with a quasi-local energy. In 
spherical symmetry, this is the Misner-Sharp-Hernandez energy \cite{MSH1, 
MSH2}, to which the more general Hawking-Hayward quasilocal energy 
\cite{Hawking68, Hayward94} reduces. In spherical symmetry, the 
Hawking-Hayward/Misner-Sharp-Hernandez quasilocal energy is in practice 
closely related to the notion of apparent horizon.

Let us begin by reviewing the concepts of null geodesic congruence and 
trapped surface necessary to define apparent horizons \cite{Wald, 
Poisson}. Consider a 
congruence of null geodesics, each with tangent $l^a=dx^a/d\lambda $, 
where $\lambda$ is an affine parameter along the geodesic. (The 
assumption of an affine parameter is not mandatory, but it simplifies 
several of the following equations.) With affine parametrization, the  
tangent  satisfies $l_a l^a=0 $ and $ l^c\nabla_c l^a=0 $.  The metric $ 
h_{ab}$ in 
the 2-space orthogonal to $l^a$ is determined as follows. Pick another 
null vector field $s^a$ such that $ s_c s^c=0 $ and 
$l^c s_c=-1$, then
\be 
h_{ab} \equiv g_{ab}+l_a s_b +l_b s_a  \,.
\ee
The tensor $ h_{ab}$ is purely spatial and ${h^a}_b$ is the 
projection operator on the 2-space orthogonal to $l^a$.
 The choice of $s^a$ is not unique, but the geometric quantities
of interest to us do not depend on it once $l^a$ is fixed \cite{Poisson}. 
Let $\eta^a$ be the geodesic deviation vector and define the tensor 
\be
B_{ab} \equiv \nabla_b \, \eta_a 
\ee
orthogonal to the null geodesics. The transverse part of the deviation 
vector is 
\be
\tilde{\eta}^a \equiv {h^a}_b \, \eta^b=\eta^a+( s^c\eta_c) 
l^a 
\ee
and the orthogonal component of $l^c\nabla_c \eta^a$, denoted by a 
tilde, is 
\be 
\widetilde{ \left( l^c \nabla_c \eta^a \right)}= {h^a}_b 
{h^c}_d {B^b}_c \, \tilde{\eta}^d \equiv { \tilde{ B^a }}_d \, 
\tilde{\eta}^d  \,.
\ee
Now decompose the transverse tensor $ \widetilde{B}_{ab}$ 
as 
\be 
\widetilde{B}_{ab}=\widetilde{B}_{(ab)} + \widetilde{B}_{[ab]}= 
\left( \frac{\theta}{2} \, h_{ab}+\sigma_{ab} \right) + \omega_{ab} \,, 
\ee
where the expansion $ \theta =\nabla_c \, l^c 
$ propagates  according to the Raychaudhuri 
equation \cite{Wald, Poisson}
\be
\frac{d\theta}{d\lambda} = 
-\frac{\theta^2}{2} -\sigma^2+\omega^2 -R_{ab}l^a l^b \,.
\ee
If the null geodesic congruence  is 
not affinely parametrized, the geodesic equation is 
\be
l^c \nabla_c l^a = \kappa \, l^a   \label{notaffineexpansion}
\ee
(where $\kappa$  is sometimes 
used as a possible notion of surface gravity) and  
\be
\theta=\nabla_c \, l^c- \kappa \,,
\ee
or 
\be
\theta_{l} = h^{ab}\nabla_{a}l_{b} = \left[ g^{ab} + \frac{l^ a s^b 
+ s^a l^b}{\left( -s^{c}l^{d} g_{cd}\right)} \right] 
\nabla_{a} l_{b}  \,.
\ee 
The difference between Eq.~(\ref{notaffineexpansion}) and our 
previous definition $\theta \equiv \nabla^c l_c$ is due to the fact that 
the geodesics in Eq.~(\ref{notaffineexpansion}) are not required to be 
affinely parametrized, while the definition   $\theta \equiv \nabla^c l_c$ 
does require an affine parameter and $\kappa$ measures precisely the 
failure of geodesics to be affinely parameterized. Since $\kappa$ 
arises purely from the choice of parametrization, its significance as a 
surface gravity is questionable and, indeed, several alternative 
definitions of ``surface gravity'' exist in the literature (see 
Refs.~\cite{NielsenYoon2008, PielhanKunstatterNielsen2011}) for reviews). 

With non-affine parametrization, the Raychaudhuri equation then 
becomes
\be
\frac{d\theta}{d\lambda} =\kappa \, 
\theta-\frac{\theta^2}{2} 
-\sigma^2+\omega^2 -R_{ab}l^a l^b \,.
\ee

A compact and orientable surface has two independent  
directions orthogonal to it, corresponding to ingoing and 
outgoing  null geodesics with tangents  $l^a$ and $n^a$, respectively.    
Let us give some basic definitions for closed 2-surfaces.\\\\
\noindent {\em Definition:} A {\it normal surface} corresponds to 
$\theta_{l} > 0$ and $\theta_{n} < 
0$.\\\\
\noindent {\em Definition:}  A {\it trapped} surface corresponds to 
$\theta_{l}<0$ and $\theta_{n}<0$.

The outgoing, in addition to the ingoing, future-directed null rays 
converge here instead of diverging and outward-propagating light is 
dragged back by strong gravity.\\\\
\noindent {\em Definition:}  A {\em marginally outer trapped} (or 
marginal) {\em surface} (MOTS) 
corresponds to  $\theta_{l} =0$ (where $l^a$ is the outgoing null normal 
to the surface) 
and $\theta_{n} < 0$.\\\\
\noindent {\em Definition:}  An {\em untrapped surface} is one with 
$\theta_{l} \theta_{n} < 0$.\\\\
\noindent {\em Definition:} An {\em antitrapped surface} corresponds to 
$\theta_{l} >0$ and $\theta_{n} > 0$.\\\\
\noindent {\em Definition:}  A {\em marginally outer trapped tube} (MOTT) 
is a 3-dimensional surface which can be foliated entirely by marginally 
outer trapped (2-dimensional) surfaces.\\

A 1965 result by Penrose \cite{Penrose65} states that if a spacetime 
contains a trapped surface, the null energy condition holds, and there is 
a non-compact Cauchy surface for the spacetime, then this spacetime 
contains a singularity. Timelike naked singularities are believed to 
be unphysical since they spoil the initial value problem and cause 
the loss of predictability. Therefore, black hole horizons assume a 
crucial role in physics, which is hiding these singularities from the 
world outside the horizon. 

Trapped surfaces are essential features in the concept of black hole. 
``Horizons'' of practical utility are identified with boundaries of 
spacetime regions containing trapped surfaces.\footnote{The mathematical 
conditions for the existence and uniqueness of marginally outer trapped 
surfaces are not completely clear.}

\subsection{Event horizons} 

\noindent {\em Definition:}  An {\em event horizon} is a connected 
component of the boundary $\partial 
\left( J^{-}( \mathscr{I}^+ )\right)$ of the causal past $ J^{-}( 
\mathscr{I}^+ )$ of future null infinity $\mathscr{I}^+ $.\\

An event horizon is a causal boundary separating a region from which 
nothing can come out to reach a distant observer from a region in which 
signals can be sent out and eventually arrive to this observer. An event 
horizon is generated by the null geodesics which fail to reach infinity. 
Provided that it is smooth, an event horizon is a null hypersurface.

In order to define and locate an event horizon, one must know all the 
future history of spacetime: the event horizon is a globally defined 
concept and has teleological nature. As an example, consider a 
collapsing Vaidya spacetime described by the geometry 
\cite{Vaidyaoriginal}
\begin{equation}
ds^2 = -\left( 1-\frac{ 2 m(v)}{r} \right) dv^2 +2dvdr+r^2 d\Omega_{(2)}^2 
\,,
\ee
where $v$ is a retarded time, $r$ is the areal radius, 
$d\Omega_{(2)}^2 \equiv d\theta^2 +\sin^2 \theta \, d\varphi^2$ is the 
line element on the unit 2-sphere, and $m(v)$ is a 
time- and radius-dependent mass function. The apparent horizon is located 
by the root of the equation $\nabla^c r \nabla_c r = g^{rr}=0$, which 
yields $r_{AH}=2m(v)$. The apparent horizon is distinct from the event 
horizon, which is given {\em approximately} by  the expression 
\cite{Balbinot, 
Booth, BenDov07, NielsenVaidya, ZhouShu}
\begin{equation}
r_{EH} = 2m \left[ 1+4\dot{m}+32 \dot{m}^2 + \mbox{O}(\dot{m}^3 ) \right] 
\,,
\end{equation}
where $\dot{m} \equiv dm/dv$. The phenomenology of the event horizons is 
reported in several studies ({\em e.g.}, \cite{Balbinot, Booth, 
BenDov07, NielsenVaidya, ZhouShu} and we will not repeat the analysis 
here. To summarize the situation, an 
event horizon forms and grows 
starting from the centre and 
an observer can cross it and be unaware of it even though his or her 
causal past consists entirely of a portion of Minkowski space. This event 
horizon cannot be detected by this observer with a physical experiment and 
it ``knows'' about events belonging to a spacetime region very far away 
and 
in its future, but not causally connected to it
 (a property called ``clarvoyance'') 
\cite{AshtekarKrishnan, BenDov07, Bengtsson11}.

Properly speaking, an event horizon ${\cal H}$ is a tube in spacetime. 
There is some language abuse in the literature, which consists of 
referring to the intersections of ${\cal H}$ with surfaces of constant 
time (which produce 2-surfaces) as ``event horizons''.

\subsection{Apparent and trapping horizons}

\noindent {\em Definition:}  A {\em future apparent horizon} is the 
closure of a  3-surface which is foliated by marginal surfaces.\\

An apparent horizon 
is defined by the  conditions on the time slicings \cite{Hayward93}  
\begin{eqnarray} 
& \theta_{l} = 0 \,,&  \label{thetal} \\ 
&&\nonumber\\
& \theta_{n} < 0 \,,& \label{thetan}  
\end{eqnarray} 
where $\theta_l$ and $\theta_n$ are the expansions of the future-directed 
outgoing and ingoing null geodesic congruences, respectively, which  
have tangent\footnote{Correspondingly, the expansions of the 
outgoing and ingoing null geodesic congruences are labelled $\theta_l$ and 
$\theta_n$.} fields $l^a$ and 
$n^a$ (outgoing 
null rays momentarily stop expanding and turn around at the horizon). 
In simple words, an event horizon is a surface from the interior of which 
nothing will ever emerge, while an apparent horizon is a surface from the 
interior of which nothing can escape now. The inequality~(\ref{thetan})  
distinguishes between black holes and white holes.

Apparent horizons are defined {\em quasi-locally} and they do not require 
the knowledge of the global structure of spacetime, including future null 
infinity. In this sense, they are much more practical entities than event 
horizons. However, apparent horizons have a serious limitation:  they 
depend on the choice of the foliation of the 3-surface with marginal 
surfaces.  This foliation-dependence is exemplified dramatically by the 
fact that non-symmetric slicings of the Schwarzschild spacetime exist for 
which there is no apparent horizon \cite{WaldIyer91, SchnetterKrishnan06}. 
In non-stationary situations, apparent horizons and event horizons do not 
coincide. This foliation-dependence problem is overlooked: it implies that 
the very existence of a dynamical black hole depends on the observer. This 
problem is alleviated (or, for purely practical purposes, largely 
eliminated), but not solved, in spherical symmetry. In this 
case, the apparent horizons coincide in all {\em spherical} foliations 
\cite{VFEllisFirouzjaeeHelouMusco17}.

In GR, a black hole apparent horizon lies inside the event horizon 
provided that the null curvature condition $R_{ab}\, l^al^b \geq 0 $ for 
all null vectors $l^a$ is satisfied. However, Hawking radiation itself 
violates the weak and the null energy conditions, as do quantum matter and 
non-minimally coupled scalars, hence apparent horizons cannot always be 
expected to lie inside event horizons in the presence of quantum fields 
or in theories of gravity alternative to GR.\\\\
\noindent {\em Definition:}  A {\em future outer trapping horizon} (FOTH) 
is the closure of a 
surface (usually a 3-surface) foliated by marginal surfaces such 
that on its 2-dimensional ``time slicings'' \cite{Hayward93}
\begin{eqnarray} 
& \theta_{l} = 0 \,,&  \\ 
&&\nonumber\\ 
& \theta_{n} < 0 \,,& \\ 
&&\nonumber\\ 
& {\cal L}_n \, \theta_{l} 
= n^{a}\nabla_{a} \, \theta_{l} < 0  \,. 
& \label{THcondition3} 
\end{eqnarray} 

The last condition distinguishes between inner and outer horizons and 
between apparent and trapping horizons (the sign distinguishes between 
future and past horizons).\\\\
\noindent {\em Definition:}  A {\em past inner trapping horizon} (PITH) is 
defined by exchanging  
$ l^a $ with $n^a $ and reversing the signs of the 
inequalities, 
\begin{eqnarray} 
& \theta_{n} = 0 \,,& \label{PITHcondition1} \\ 
&&\nonumber\\ 
& \theta_{l} > 0 \,,& \label{PITHcondition2} \\ 
&&\nonumber\\ 
& {\cal L}_l \theta_n = 
l^{a}\nabla_{a} \, \theta_{n} > 0 \,.& \label{PITHcondition3} 
\end{eqnarray}

The PITH identifies a white hole or a cosmological horizon.  As one moves 
just inside an outer trapping horizon, one encounters trapped surfaces, 
while trapped surfaces are encountered as as one moves just outside an 
inner trapping horizon. As an example, consider the Reissner-Nordstr\"om 
black hole with the natural spherically symmetric foliation. The event 
horizon $r=r_{+}$ is a future outer trapping horizon (FOTH), the inner 
(Cauchy) horizon $r=r_{-}$ is a future inner trapping horizon (FITH), 
while the white hole horizons are past trapping horizons (PTHs).

Black hole trapping horizons have been associated with thermodynamics. It 
is claimed that it is the trapping horizon area, and not the area of the 
event horizon, which should be associated with entropy in black hole 
thermodynamics \cite{Haijcek87, Hiscock89, Collins92, Nielsen}. This claim 
is somewhat controversial \cite{Sorkin97, CorichiSudarsky02, 
NielsenFirouzjaee12}. The Parikh-Wilczek \cite{ParikhWilczek00} 
``tunneling'' approach, originally applied to study Hawking radiation 
from stationary black holes, is in principle applicable 
also to apparent and trapping horizons.

In general, trapping horizons do not coincide with event 
horizons. Dramatic examples are spacetimes which possess 
trapping horizons but not event horizons \cite{RomanBergmann83, 
Hayward06}.

\subsection{Spherical symmetry}

As usual, assuming a symmetry greatly simplifies a physical problem, and 
this is even more so for the assumption of spherical symmetry when 
discussing apparent horizons and dynamical black holes. The notion of 
internal energy in black hole thermodynamics is crucial for the first law 
and, in spherical symmetry, it is universally identified with 
the Misner-Sharp-Hernandez mass notion \cite{MSH1, MSH2}.

The Misner-Sharp-Hernandez mass \cite{MSH1, MSH2} is defined in 
GR\footnote{The  Misner-Sharp-Hernandez mass is also well defined in 
Gauss-Bonnet gravity \cite{HidekiLovelock, Hideki2}.} and in spherical 
symmetry. The more general Hawking-Hayward quasilocal 
energy  \cite{Hawking68, Hayward94} reduces to the Misner-Sharp-Hernandez 
notion in spherical symmetry. A spherically symmetric line element can be 
written as 
\be
ds^2=h_{ab}dx^a dx^b +R^2 d\Omega_{(2)}^2  \;\;\;\;\;\;\; (a,b=1,2)
\ee
where $R$ is the areal radius, a geometrical quantity defined in a 
coordinate-invariant way using the area ${\cal A}$ of 2-spheres of 
symmetry, 
$R=\sqrt{{\cal A}/(4\pi)}$, and $h_{ab}$ is the 2-metric on the 
submanifold 
orthogonal to the time and radial directions. 
Then the Misner-Sharp-Hernandez mass $M_{MSH}$ is defined in an 
invariant way by the scalar equation \cite{MSH1, MSH2}
\be
1-\frac{2M_{MSH}}{R} \equiv \nabla^c R  \, \nabla_c R \,. 
\label{AHequation}
\ee
It is often useful to use the formalism of Nielsen and  Visser 
\cite{NielsenVisser06}, in which a  general spherical line element is 
written as 
\be
ds^2=-\mbox{e}^{-2\phi (t, R)} \left[ 1-\frac{2M(t,R)}{R} 
\right] dt^2 +\frac{dR^2}{1-\frac{2M(t,R)}{R} } 
+R^2d\Omega_{(2)}^2 \,, \label{NielsenVisser} 
\ee
where $M(t,R)$, {\em a 
posteriori}, turns out to be the  Misner-Sharp-Hernandez 
mass. This notation interpolates between a gauge originally used to 
describe wormholes and the more familiar Schwarzschild-like coordinates.

The line element~(\ref{NielsenVisser}) is recast  in 
Painlev\'e-Gullstrand coordinates as 
\be
ds^2=- \frac{ \mbox{e}^{-2\phi } }{ \left( \partial\tau 
/\partial t\right)^2} \left( 1-\frac{2M}{R} \right) 
d\tau^2 +  \frac{ 2\mbox{e}^{-\phi}}{ \partial\tau/\partial t } 
\sqrt{ \frac{2M}{R}} \, d\tau dR 
+dR^2   +R^2d\Omega_{(2)}^2 \,, 
\ee
where the functions $\phi( \tau,R)$ and $M(\tau, R)$ are defined 
implicitly. By introducing the quantities 
\begin{eqnarray}
c \left(\tau, R \right) & \equiv & \frac{ \mbox{e}^{-\phi(t,R)} 
}{\left( \partial\tau/ \partial t \right)}  \,,\\
&&\nonumber\\
v \left(\tau, R \right) & \equiv & 
\sqrt{\frac{2M(t,R)}{R} } \, \frac{ \mbox{e}^{-\phi(t,R)} 
}{\partial\tau/\partial t} =c \,  
\sqrt{\frac{2M}{R} }   \,,
\end{eqnarray}
the line element becomes \cite{NielsenVisser06}
\be
ds^2= -\left[ c^2\left(\tau, R \right) - v^2\left(\tau, R \right) 
\right] d\tau^2 +2v \left(\tau, R \right) d\tau dR +dR^2 +R^2 
d\Omega_{(2)}^2 \,.
\ee
In this gauge, the outgoing radial null geodesics have tangents   
\be
l^{\mu}=\frac{1}{c(\tau, R)} \Bigg( 1, c(\tau, R)-v(\tau, R), 0,0 
\Bigg) \,,
\ee
while the ingoing radial null geodesics have tangents
\be
n^{\mu}=\frac{1}{c(\tau, R)} \Bigg( 1, -c(\tau, R)-v(\tau, R), 
0,0 \Bigg) 
\ee
with 
\be
g_{ab}l^a n^b=-2 \,,
\ee
and the expansions of the radial null geodesic congruences are 
\be
\theta_{l,n} = \pm \frac{2}{R} \left( 1 \mp \sqrt{ \frac{2M}{R} } \right) 
\,.
\ee
Then a  sphere of radius $R$ is trapped if $R<2M $, marginal if $R=2M $,   
untrapped if $R>2M $. The apparent horizons are located by
\be
\frac{2M\left( \tau, R_{AH}\right)}{R_{AH}(\tau)}=1
 \;\; \Longleftrightarrow \;\;  \nabla^c 
R\nabla_c R \left. \right|_{AH} =0 \Longleftrightarrow \;\; 
 g^{RR} \left. \right|_{AH}=0 \,.    
\ee
The inverse metric is
\be
\left( g^{\mu\nu} \right)=\frac{1}{c^2} \left(
\begin{array}{cccc}
-1 & v & 0 & 0 \\
&&&\\
v & (c^2-v^2) &0 &0 \\
&&&\\
0 &0 & 1/R^2 & 0\\
&&&\\
0 & 0 & 0 & 1/(R^2 \sin^2 \theta )
\end{array} \right) \,.
\ee
In practice, the condition $g^{RR}=0$ is a very convenient 
recipe  to locate the apparent horizons in  
spherical symmetry. Of course, one does not need to use a particular 
gauge, or to use the areal radius as a radial coordinate: the apparent 
horizons are still located by Eq.~(\ref{AHequation}) (with the 
ever-present {\em caveat} 
that apparent horizons are foliation-dependent if non-spherical foliations 
are used).

The normal vector  
$\nabla_a R$ to the surfaces $R=$const. becomes null on the 
apparent horizons. Moreover, in the Nielsen-Visser gauge we have 
\cite{NielsenVisser06} 
\be
{\cal L}_{n} \theta_l \left. \right|_{AH}=
- \frac{2 \left(1-2M'_{AH} \right)}{R_{AH}^2} \left( 
1+\frac{\dot{R}_{AH} }{2c_{AH}} \right) \,,
\ee
where $'=\partial/\partial R$ and $\dot{} \equiv \partial/\partial \tau$.
If $1-2M'_{AH}>0 $, the  horizon is outer. The condition for the apparent 
horizon to be also a trapping horizon  is 
\be
 \dot{R}_{AH} >-2c_{AH} \,.
\ee
If matter satisfies the null energy condition, and assuming the 
Einstein equations, the area of the apparent horizon cannot 
decrease. The Nielsen-Visser  surface gravity is 
\be
\kappa_l ( \tau ) = \frac{ 1-2M' \left( \tau, R_H( \tau) 
\right) }{ 2 R_H( \tau) }  \,.
\ee

\section{A selection of exact solutions in various theories of gravity}
\label{sec:3}

Having introduced the basic quantities, we can now proceed to see a little 
bestiary of analytical solutions of the field equations of 
various theories of gravity describing dynamical black holes embedded in 
cosmological spacetimes at least part of the time. Not many such exact 
solutions are known and 
some of them have physical problems such as negative energy densities, 
usually in spatial regions close to a black hole apparent horizon. 
Nevertheless, a complete review would be too long\footnote{See 
\cite{mybook} for details.} and a somewhat arbitrary selection 
needs to be made, but the geometries appearing most often in the 
literature are reported below. For extra discussion of cosmological black 
holes in alternative gravity see \cite{Sotiriou, HerdeiroRadu, mybook, 
TretyakovaLatosh}.

\subsection{Schwarzschild-de Sitter/Kottler spacetime} 

The oldest known cosmological black hole geometry is the 
Schwarzschild-de Sitter/Kottler solution of GR and of 
many other theories \cite{Kottler1918}. It solves the vacuum 
Einstein equations with positive cosmological constant $\Lambda$
\begin{equation}
R_{ab}-\frac{1}{2}\, g_{ab} {\cal R}=-\Lambda g_{ab} 
\,,\label{efelambda}
\ee
where $R_{ab}$ is the Ricci tensor and ${\cal R} \equiv g^{ab}R_{ab}$ is 
its trace. The Jebsen-Birkhoff theorem familiar from GR textbooks and 
stating the 
uniqueness of the Schwarzschild solution under the assumptions of 
vacuum, spherical symmetry, and asymptotic flatness can easily be 
generalized to 
show that the Schwarzschild-de Sitter/Kottler geometry is the unique 
spherically symmetric solution of Eq.~(\ref{efelambda}) with $\Lambda>0$ 
\cite{Synge, SchleichWitt, FaraoniCardiniChung}. The Schwarzschild-de 
Sitter/Kottler line element is 
\begin{eqnarray}
ds^2 &=& -\left( 
1-\frac{2m}{R}-H^2R^2 \right) dt^2+ \left( 1
-\frac{2m}{R}-H^2R^2 \right)^{-1} 
dR^2  \\
&&\nonumber\\
&\, & + R^2 d \Omega_{(2)}^2
\end{eqnarray} 
in static coordinates, where $H=\sqrt{\Lambda/3}$. This geometry is 
locally static in the region 
comprised between the black hole and the cosmological horizons  $ R_1 <R 
<R_2 $ (with $R_{1,2}$ defined below), where a timelike Killing vector 
exists. The apparent horizons are  
located by the usual equation $g^{RR} =1-\frac{2m}{R}-H^2R^2=0 $. 
The formal roots of this 
cubic equation are 
\begin{eqnarray} 
R_1 &=&\frac{2}{\sqrt{3}H}\sin\psi \,,\\ 
&&\nonumber\\ 
R_2 &=&\frac{1}{H}\cos\psi 
-\frac{1}{\sqrt{3}H}\sin\psi \,,\\ 
&&\nonumber\\ 
R_3&=&-\frac{1}{H}\cos\psi -\frac{1}{\sqrt{3}H} \sin\psi \,, 
\end{eqnarray} 
with $\sin (3\psi )=3\sqrt{3} \, mH$. Here    
$m$ and $ H$ are positive, which implies that $ R_3<0$ and there 
are at most two apparent horizons. When $R_1$ and $R_2$ are 
real, $R_1$ is a black hole apparent horizon while $R_2$ is a cosmological 
apparent horizon. Both apparent horizons are null surfaces, due to the 
(locally) static character of this geometry. Three situations are 
possible:

\begin{itemize}

\item Two apparent horizons exist if $ 0<\sin( 3\psi ) < 1$.

\item If $\sin (3\psi)=1$  the apparent horizons  
coincide, giving the extremal Nariai black hole.

\item For $\sin (3\psi ) >1$ there is a naked singularity. The 
physical interpretation of this situation is that the 
black hole horizon becomes larger than  the cosmological one and 
effectively disappears so, in the region below the cosmological horizon,  
the central singularity is  not screened by a black hole horizon.

\end{itemize}

\subsection{McVittie solution}

The McVittie solution of the Einstein equations \cite{McVittie33}  
\begin{equation}
R_{ab}-\frac{1}{2} \, g_{ab} {\cal R}=8\pi T_{ab}  \label{efe}
\end{equation}
generalizes the Schwarzschild-de Sitter/Kottler geometry and it contains 
it as a  special case. The McVittie spacetime represents a central object 
embedded in a FLRW universe. Many papers over the past eighty years have 
focused on this GR solution. There 
are also also versions of McVittie with negative cosmological constant 
and/or electric charge. Here we focus on the uncharged McVittie 
metric 
with spatially flat FLRW ``background''.

The original motivation for the McVittie solution was the study of   
the cosmological expansion (which in those days was still a fairly recent 
discovery) on local systems. This problem was later investigated also by 
Einstein, who was unaware of McVittie's work and led to the 
Einstein-Straus vacuole, or Swiss-cheese model \cite{EinsteinStraus1, 
EinsteinStraus2}.  
McVittie made a crucial simplifying assumption in the derivation of his 
solution using spherical coordinates \cite{McVittie33}: the ``no-accretion 
condition'' $G_0^1=0$. This implies $T_0^1=0$ due to the Einstein 
equations and, therefore, zero radial energy flow onto, or out of, the 
central inhomogeneity. The McVittie line element in isotropic coordinates 
is 
\cite{McVittie33} 
\be
ds^2=-\frac{  \left(1-\frac{m(t)}{2\bar{r}} \right)^2}{
\left(1+\frac{m(t)}{2\bar{r}} \right)^2} \, dt^2+
a^2(t) \left( 1+\frac{m(t)}{2\bar{r}} \right)^4 \left( 
d\bar{r}^2 +\bar{r}^2 d\Omega_{(2)}^2 \right) \,,
\ee
and the McVittie no-accretion condition $G_0^1=0$ translates to 
\be
\frac{\dot{m}}{m}+\frac{\dot{a}}{a}=0 \,,
\ee
with solution 
\be
m(t)=\frac{m_0}{a(t)} \,,\;\;\; \;\;\;\;\;m_0 = \mbox{const.}
\ee
The McVittie line element reduces to the Schwarzschild one if $ a \equiv 
1$ and to the FLRW one if $m = 0$. There are spacetime singularities   
at  $\bar{r}=m/2$ and $\bar{r}=0$. The singularity at finite radius  
$\bar{r}=m/2$ effectively separates two causally disconnected spacetimes 
and only the region  $\bar{r}> m/2$ is usually considered. In this region 
the  energy density $\rho(t)$ of the 
matter fluid (which depends only on time) is finite, but the pressure  
\be
P \left( t, \bar{r} \right) 
=-\, \frac{1}{8\pi} \left[ 3H^2+\frac{2\dot{H}\left( 
1+\frac{m}{2\bar{r}} \right)  }{1-\frac{m}{2\bar{r}} }\right] 
\ee
(which depends on both time and radius) diverges  
as $\bar{r} \rightarrow m/2$, together with the Ricci scalar 
\be
{\cal R} =8\pi \left( 3P-\rho \right) \,,
\ee
except in the case of a  de Sitter ``background'' with $\dot{H}=0$.

The apparent horizons of the McVittie geometry have been studied in  
\cite{Nolan1, Nolan2, Nolan3, LiWang06, ZambranoNandra}. Rewrite the 
metric using the 
areal radius 
\be
R (t, \bar{r}) \equiv a(t) \bar{r} \left( 1+\frac{m}{2\bar{r}} \right)^2 
\,,
\ee
which leads to
\be
ds^2 = -\left( 1-\frac{2m_0}{R}-H^2R^2 \right) dt^2
+\frac{dR^2 }{1-\frac{2m_0}{R} } 
-\frac{2HR \, dtdR }{\sqrt{1-\frac{2m_0}{R} } } +R^2 
d\Omega_{(2)}^2 \,.
\ee
The cross-term in $dtdR$ can be eliminated defining the new time 
coordinate  $T( t, R)$ by
\be
dT=\frac{1}{F} \left( dt+\beta dR \right)  \,,
\ee
where $F(t, R)$ is an integrating factor. The cross-term in the new 
coordinate system vanishes by imposing    
\be
\beta (t, R) = 
\frac{HR}{  \sqrt{1-\frac{2m_0}{R}}\, 
  \left( 1-\frac{2m_0}{R}-H^2R^2 \right) } \,.
\ee
The line element is then recast as 
\be
ds^2 = -\left( 1-\frac{2m_0}{R}-H^2R^2 \right) F^2 
dT^2 + \frac{dR^2 }{  1-\frac{2m_0 }{R}-H^2R^2  } 
+R^2  d\Omega_{(2)}^2 \,,
\ee
where $ \bar{r}=m/2 $ corresponds to $ R= 2m \, 
a(t)= 2 m_0$, that is, the finite radius singularity does not expand with 
the rest of the universe in which it is embedded.

The McVittie metric admits arbitrary FLRW ``backgrounds'' 
generated by fluids with  any constant equation of state. Let us restrict   
to a fluid which reduces to dust at spatial infinity ($w=0$), for the sake 
of 
simplicity.  The fluid pressure is
\be
P\left(t,R \right)=\rho(t) \left( 
\frac{1}{\sqrt{1-\frac{2m}{R}}} -1 \right) \,.
\ee
The apparent horizons are located at the roots of the equation  
\be
g^{RR}=1-\frac{2m}{R}-H^2(t) \, R^2=0 \,.
\ee
This equation is the same cubic already seen in the  
Schwarzschild-de Sitter/Kottler case, but now it has a  time-dependent 
coefficient $H(t)$. Its roots  $R_{1,2}(t)$ are given again by the 
expression already seen, but now  with time-dependent coefficient  $H(t)$. 
This means that the location of the apparent horizons 
depends on time ({\em i.e.}, on the comoving time $t$ of the FLRW 
``background''). Both 
apparent horizons exist if $mH(t)<1/(3\sqrt{3})$, an inequality which is 
satisfied only if $t>t_*$.  The critical time 
at which $mH(t)=1/(3\sqrt{3})$ for a dust ``background'' is  
\be
t_* =2\sqrt{3} \, m \,.
\ee
There are three distinct epochs in the history of this inhomogeneous 
universe (Fig.~\ref{fig:McVittieAHs}): 

\begin{enumerate}

\item  For $t<t_*$, it is $m>\frac{1}{3\sqrt{3} 
\,H(t)}$ and both $R_1(t)$ and $R_2(t)$ are complex. 
There are no apparent horizons.

\item The critical time $t=t_*$ corresponds to 
$m=\frac{1}{3\sqrt{3}\,H(t)}$.  The two roots $R_{1,2}(t)$ coincide at a 
real value,  and there exists a single apparent horizon at 
$R_*=\frac{1}{\sqrt{3}\,H(t_*)}$.
 
\item For $t>t_*$, it is $ m < 
\frac{1}{3\sqrt{3}\,H(t)}$, and there are two 
apparent horizons at real radii $R_{1,2}(t)>0$.

\end{enumerate}

\begin{figure}
\centering
\includegraphics[width=11 cm]{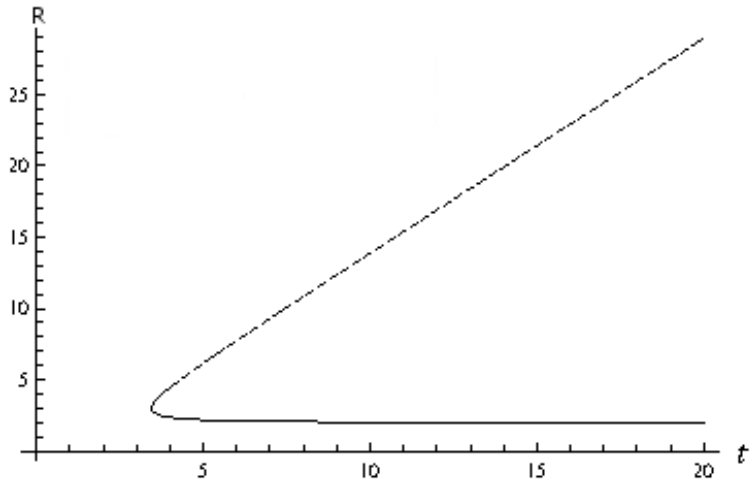}
\caption{The McVittie cosmological (dashed) and black hole (solid) 
apparent horizons in a dust-dominated ``background'' universe. Time $t$ 
and radius $R$ are in units of $m$.
\label{fig:McVittieAHs}} 
\end{figure}

The Hubble function $H(t)$ diverges near the Big Bang, 
when the mass coefficient $m$ stays supercritical at
 $m>\frac{1}{3\sqrt{3}\,H(t)}$.  A black hole apparent horizon cannot 
be accommodated in this small universe and, at $t<t_{*}$ there is a  
naked singularity at $R=2m_0$. At $t_{*}$ an instantaneous black hole 
apparent horizon and a  cosmological apparent horizon appear together at 
areal radius $R_1(t_*)=R_2(t_*) =\frac{1}{\sqrt{3}\,H(t_*)}$, in analogy 
with the extremal Nariai black hole. 
As  $t>t_{*}$, this single horizon splits 
into an evolving black hole apparent horizon surrounded by 
an evolving cosmological horizon.   The black hole apparent horizon shrinks, 
asymptoting to the  $2m_0$ singularity  as $t\rightarrow +\infty$.

Motivated by the discovery of dark energy, and by recurrent claims by 
astronomers that this dark energy may have a phantom equation of state 
with $w \equiv P/\rho  <- 1$ (although such a  form of dark energy 
seems to be in 
extreme conflict with known theory), one can consider a phantom 
``background'' universe in the McVittie solution. In this case the FLRW 
cosmic scale factor is
\be
a(t)=\frac{A}{ \left( t_{rip}-t 
\right)^{\frac{2}{3| w+1|} }} \,, \;\;\;\;\;  
H(t)=\frac{2}{3|w+1|} \, \frac{1}{t_{rip}-t} \,,
\ee
where $t_{rip}$ is the time of the Big Rip and $A$ is a positive constant. 
In this situation, the 
behaviour of the apparent horizons (Fig.~\ref{fig:phantomMcV}) appears to 
be the time reversal of the behaviour already seen in a non-phantom 
universe.

\begin{figure}
\centering
\includegraphics[width=11 cm]{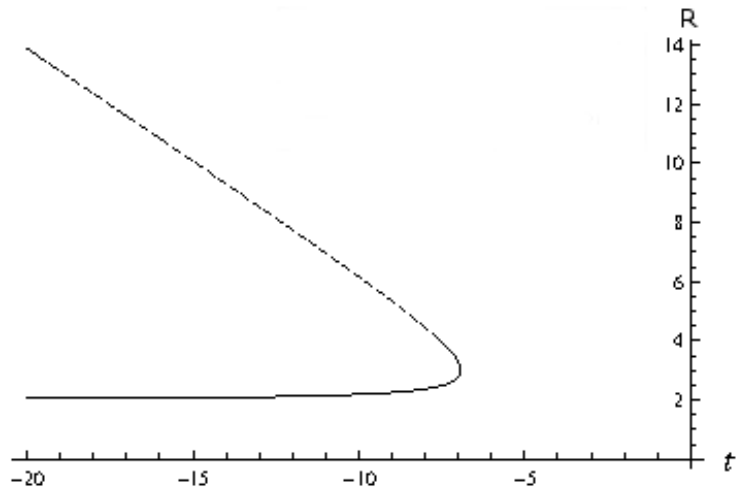}
\caption{\label{McV2} The 
McVittie apparent  horizons  in a 
phantom-dominated universe (here $w=-1.5$ and $t_{rip}=0$).
\label{fig:phantomMcV}}
\end{figure}

An idealized interior solution for the McVittie exterior metric 
has been found \cite{NIS}. It  describes 
a relativistic star of uniform density in a FLRW ``background'' 
\cite{NIS}. The equivalent of the Tolman-Oppenheimer-Volkov  equation 
in the interior of a McVittie star is derived in \cite{VFJacques08}.

Recent works on the McVittie spacetime have studied its conformal 
structure 
\cite{Klebanetal, LakeAbdelqader11, Lake2, daSilvaetal12},  which 
entails  
integrating numerically the null geodesics or deriving general 
analytical results upon making some assumptions about the cosmic 
expansion.  Lake \& 
Abdelqader \cite{LakeAbdelqader11} have found that  null 
geodesics asymptote to the singularity without entering it. Depending on 
the form of the scale factor, a bifurcation surface may appear which 
splits the spacetime boundary into a black hole horizon in the future 
and a white hole horizon in the past (it is not clear whether this is 
due directly to the McVittie no-accretion condition). 
da Silva, Fontanini, and Guariento \cite{daSilvaetal12}   find that 
the presence of this white hole horizon depends crucially on the expansion 
history of the universe. 

It was found recently that the McVittie geometry is also a 
solution of cuscuton theory, a special 
Ho\u{r}ava-Lifschitz theory, and of shape dynamics 
\cite{Abdallaetal14,  
Gomesetal11, GuarientoMercati}.

\subsubsection{Generalized McVittie solutions}

It is interesting to remove the no-accretion restriction from the 
 McVittie solutions  \cite{VFJacques08}. In principle, this becomes the  
``Synge approach'' in which one postulates a metric and runs the 
Einstein equations $R_{ab}-{\cal R} g_{ab}/2=8\pi T_{ab}$ from the 
left to the right to derive the form of an effective stress-energy 
tensor $T_{ab}$  that generates this metric. Usually this approach is 
fruitless 
because the effective $T_{ab}$ thus generated is physically pathological 
and violates all the energy conditions. However, reasonable 
matter sources exist in this case. Assume the line element to be 
\be 
ds^2= -\frac{B^2\left(t, \bar{r} \right)}{
A^2\left(t, \bar{r} \right)}\, dt^2 +a^2(t) A^4 \left(t, \bar{r} 
\right) \left( d\bar{r}^2+\bar{r}^2 d\Omega^2_{(2)} \right) \,,
\ee
where 
\begin{eqnarray}
m(t) & \geq &  0 \,, \\
&&\nonumber\\
A \left(t, \bar{r} \right) &=& 1+\frac{m(t)}{2\bar{r}} \,,\\
&&\nonumber\\
B \left(t, \bar{r} \right) &=& 1-\frac{m(t)}{2\bar{r}}  \,,
\end{eqnarray} 
and where now the new function $m(t)$ is not determined by the McVittie 
no-accretion condition. The mixed Einstein tensor is 
\begin{eqnarray}
G_0^0 &=& -\, \frac{3A^2}{B^2}\left( 
\frac{\dot{a}}{a} +\frac{\dot{m}}{\bar{r}A} \right)^2 \,, \\
&&\nonumber\\
G_0^1 &=&  \frac{2m}{ \bar{r}^2 a^2 A^5 B} \left( 
\frac{\dot{m}}{m} + \frac{\dot{a}}{a} \right) \,, \\
&&\nonumber\\
G_1^1 &=& G_2^2 =G_3^3 =
- \frac{A^2}{B^2}\left\{ 
2  \frac{d}{dt} \left( 
\frac{\dot{a}}{a}+\frac{\dot{m}}{\bar{r}A} 
\right) 
+ \left( \frac{\dot{a}}{a}+\frac{\dot{m}}{\bar{r}A} \right) 
\right.\\
&&\nonumber\\
&& \left. \cdot \left[
3 \left( \frac{\dot{a}}{a}+\frac{\dot{m}}{\bar{r}A} \right)
+ \frac{2\dot{m}}{\bar{r}AB} \right]\right\}  \,.
\end{eqnarray}
For the special subclass of solutions which has  $m=m_0=$const., the 
quantity
\be
C\equiv  \frac{\dot{a}}{a}+\frac{\dot{m}}{\bar{r}A} =
\frac{\dot{M}}{M}-\frac{\dot{m}}{m} \frac{B}{A}
\ee
reduces to the familiar $ \dot{M}/M $, where  
\be
 M(t) \equiv m_0 a(t) \,;
\ee
this is the non-rotating Thakurta solution \cite{Thakurta81}, which will 
re-appear later in our discussion. 

At $\bar{r}=m/2$, the quantity $C$  reduces  to 
\be
C_{\Sigma}= \frac{\dot{a}}{a}+\frac{\dot{m}}{m} =
\frac{\dot{M}}{M}  \,.
\ee
The  McVittie solutions correspond to $C_{\Sigma}=0$ everywhere,  
while the non-rotating Thakurta solution corresponds to $C=C_{\Sigma}=H$ 
everywhere. The  Ricci scalar  
\be
{\cal R} = \frac{3A^2}{B^2}\left( 2\dot{C} 
+4C^2 +\frac{ 
2\dot{m}C}{\bar{r}AB} \right)
\ee
diverges at $\bar{r}=m/2$, unless $m$ is 
a constant.

We must now discuss the matter source. If this is a single perfect fluid, 
only the McVittie solutions are possible. However, imperfect fluids can be 
matter sources for generalized McVittie spacetimes. Two cases have been 
studied \cite{VFJacques08} and are reported below. In addition, since the 
McVittie condition has been removed and we now have a radial energy flow, 
we must model it somehow and an imperfect fluid is the simplest model 
for this purpose.

\subsubsection{Imperfect fluid and no radial mass flow}

In this case the right hand side of the Einstein 
equations~(\ref{efe}) contains the matter stress-energy tensor
\be
T_{ab}=\left( P+\rho \right)u_a u_b +P g_{ab}+ q_a u_b+q_b u_a \,, 
\label{imperfectfluid}
\ee
where the purely spatial vector $q^c$ describes a radial energy flow, 
\be
u^{\mu}=\left( \frac{A}{B}, 0,0,0 \right) \,, \;\;\;\;\;
q^{\alpha}=\left( 0, q, 0,0 \right) \;, \;\;\;\;\; q^cu_c=0 
\ee
and $ u^c u_c=-1 $.  The Einstein equations yield
\be
\frac{\dot{m}}{m}+\frac{\dot{a}}{a}= -\frac{4\pi G}{m}\, 
\bar{r}^2 a^2 A^4 B^2 q  \,.
\ee
The radial energy flow, the area ${\cal A}$ of a 2-sphere, 
and the accretion rate are related by \cite{VFJacques08}  
\be
\dot{M} (t)=- a B^2 {\cal A} q   \,.
\ee
In the case of inflow ($q<0$), on a sphere of radius $\bar{r} \gg m$ this 
accretion rate becomes $\dot{M} \simeq a {\cal A} \left| q \right| $. The 
mass $M$ changes due to inflow of matter and to the evolution of the 
cosmological fluid in it. The energy density and the pressure of the fluid 
are 
\begin{eqnarray}
\rho \left( t, \bar{r}\right) &=&
\frac{3A^2}{8\pi B^2} \left( 
\frac{\dot{a}}{a}+\frac{\dot{m}}{\bar{r}A} \right)^2 
\geq 0 \,, \\
&&\nonumber\\
P \left( t, \bar{r}\right) &=&
\frac{- A^2}{8\pi  B^2} \left\{ 
2 \frac{d}{dt} \left( \frac{\dot{a}}{a}+\frac{\dot{m}}{\bar{r}A} 
\right) \right. \nonumber\\ 
&&\nonumber\\
&\, & \left. + \left( \frac{\dot{a}}{a}  
+\frac{\dot{m}}{\bar{r}A} \right)\left[ 3
\left( \frac{\dot{a}}{a} + \frac{\dot{m}}{\bar{r}A} 
\right) + \frac{2\dot{m}}{\bar{r}AB} \right] \right\} \,. 
\end{eqnarray}
The evolution of the quantity $C$ is regulated by the generalized 
Raychaudhuri equation \cite{VFJacques08}
\be
\dot{C}=-\,\frac{3C^2}{2}-\frac{\dot{m}}{\bar{r}AB}\, C -4\pi  
\,  \frac{B^2}{A^2} \, P =-4\pi  \, \frac{B^2}{A^2}\left( P+\rho \right) 
- \frac{\dot{m}C}{\bar{r}AB} \,. 
\ee

\subsubsection{Imperfect fluid and radial mass flow}

A more general situation is the one in which both an imperfect fluid and 
a radial flow of material and energy are allowed, as described by the 
stress-energy tensor~(\ref{imperfectfluid}) with 
\be
u^{\mu}=\left( \frac{A}{B}\sqrt{1+a^2A^4u^2}, u, 0, 0  
\right)\,, \;\;\;\;\;\;\;\;\;
q^{\mu}=\left( 0,q,0,0 \right) \,,
\ee
({\em i.e.}, with non-vanishing radial velocity) and with spacelike 
flux density
\be
q=-\left(P+\rho \right)\frac{u}{2} \,. 
\ee
In this case the accretion rate onto the central object is 
\be
\dot{M} =-\frac{1}{2} aB^2 \sqrt{1+a^2A^4 u^2} \left( P+\rho 
\right){\cal A}u \,,
\ee
and the energy density is found to be  \cite{VFJacques08}
\be
8\pi  \rho= \frac{A^2}{B^2}\left[ 3C^2 +\left( 
\dot{C}+\frac{\dot{m}C}{\bar{r}AB} \right) 
\frac{2 a^2A^4u^2}{1+a^2A^4 u^2} \right] \,. 
\ee
The generalized McVittie geometry is also a solution of Horndeski theory  
\cite{Afshordietal14}.

\subsubsection{The non-rotating Thakurta solution}
 
The choice $M(t)=m_0 \, a(t)$ selects a special subclass of generalized 
McVittie solutions of the Einstein equations, the non-rotating Thakurta 
solution \cite{Thakurta81}. The ``comoving mass'' attractor family 
coincides with the non-rotating Thakurta solution \cite{Thakurta81} of GR 
discovered in 1981. This is the special case, corresponding to zero 
angular momentum, of the Thakurta solution describing a Kerr black hole 
embedded in a FLRW background \cite{Thakurta81}. Recently, the 
non-rotating Thakurta solution has been discussed, with different 
purposes, in Refs.~\cite{Culetu13, MelloMacielZanchin17}. The apparent 
horizons exhibit the same phenomenology of apparent horizons described by 
a C-shaped curve in the $\left( t, R \right)$ plane which appears in the 
McVittie spacetime and in other solutions of GR, including generalized 
McVittie solutions and some Lema\^itre-Tolman-Bondi models (keep in mind, 
however, that generalized McVittie geometries are also solutions of 
Horndeski gravity).

The non-rotating Thakurta subclass is 
not a mere curiosity, but it is important because it is 
a late-time attractor of generalized McVittie 
solutions \cite{GaoChenVFShen08}. Given that the Jebsen-Birkhoff theorem
\cite{Birkhoff1, Birkhoff2}  
fails in the presence of matter and/or when the asymptotics are not 
Minkowskian, and then  no ``general solutions'' are known, this result is 
 significant because it provides a general solution {\it in the 
generalized McVittie class}, provided that one waits long enough (either 
in a universe expanding forever or in a phantom-dominated universe 
approaching a Big Rip singularity at a finite future). The non-rotating 
Thakurta solution is generic under certain assumptions, in the sense that 
all other generalized McVittie solutions approach it at late times.

Another peculiarity of this ``comoving mass'' Thakurta attractor resides 
in the 
structure of its apparent horizons. In general, in spherical 
symmetry, the radii of the apparent horizons can 
only be found numerically because Eq.~(\ref{AHequation}) locating them is 
trascendental. It is very rare to be able to obtain analytical 
explicit expressions for apparent horizon radii, which would be useful, 
for example, for quantum field theory calculations in these curved 
spacetimes to determine Hawking radiation fluxes and so on.  This task is 
possible for the non-rotating Thakurta family of solutions, for which the  
areal radii of the cosmological and black hole apparent horizons are 
\cite{GaoChenVFShen08}
\be
R_{c,b} = {\frac{1}{2H}}{\left(1 \pm \sqrt{1-8m_0\dot{a}} \, 
\right)} \,.
\ee

\subsection{The Husain-Martinez-Nu\~nez solution}

The 1994 Husain-Martinez-Nu\~nez solution of the Einstein equations brings 
in new 
phenomenology of the apparent horizons with respect to what we have seen 
thus far. The Husain-Martinez-Nu\~nez spacetime  describes an 
inhomogeneous universe with 
a spatially flat FLRW ``background'' sourced by a 
free, minimally coupled, scalar field. 
The Einstein equations~(\ref{efe}) assume the form
\begin{equation}
R_{ab}-\frac{1}{2} \, g_{ab} {\cal R}=8\pi \left( \nabla_a \phi \nabla_b 
\phi -\frac{1}{2} g_{ab} \nabla^c \phi \nabla_c \phi \right) \,,
\end{equation}
while the free scalar field $\phi$ obeys the curved space Klein-Gordon 
equation
\begin{equation}
\Box \phi =0 \,.
\end{equation}
The Husain-Martinez-Nu\~nez line element is 
\begin{eqnarray}
ds^2 &=& \left( A_0 \eta +B_0 \right) \left[ - \left( 
1-\frac{2C}{r}\right)^{\alpha} 
d\eta^2 +\frac{dr^2}{\left( 1-\frac{2C}{r}\right)^{\alpha} 
} \right. \nonumber\\
&&\nonumber\\
&\, & \left. + r^2 \left( 
1-\frac{2C}{r}\right)^{1-\alpha} 
d\Omega_{(2)}^2 \right] \,,\\
&&\nonumber\\
\phi(\eta, r ) &=& \pm \frac{1}{4\sqrt{\pi}} \, \ln \left[ 
D\left( 1-\frac{2C}{r}\right)^{\alpha/\sqrt{3}} 
\left( A_0 \eta +B_0 \right)^{\sqrt{3}} \right] 
\end{eqnarray}
where $A_0 , B_0 , C, D\geq 0$ are constants, $\alpha =\pm \sqrt{3}/2$,  
and $\eta >0$. The additive constant $B_0$ becomes irrelevant and can be 
dropped if $A_0\neq 0$. When $A_0=0$, the Husain-Martinez-Nu\~nez line 
element   
degenerates into the  static Fisher one \cite{Fisher48} 
\be
ds^2=-V^{\nu}(r) \, d\eta^2 +\frac{dr^2}{V^{\nu}(r)} +r^2 
V^{1-\nu}(r) d\Omega_{(2)}^2  \,,
\ee
where $V(r)=1-2\mu/r$, $\mu$ and $\nu$ are parameters,  and 
the Fisher scalar field is 
\be
\psi(r)=\psi_0 \ln V(r) \,.
\ee
The Fisher solution also goes by the names  
Buchdahl-Janis-Newman-Winicour-Wyman 
solution because it was rediscovered many times by these authors 
\cite{Fisher3, Fisher4, Fisher5, Fisher6}. It 
always hosts a naked 
singularity at $r=2C$ and is asymptotically flat. The 
general Husain-Martinez-Nu\~nez metric is 
conformal to the Fisher metric with conformal 
factor 
$\Omega (\eta) =\sqrt{A_0 \eta+B_0}$ equal to the scale factor 
$a(\eta)$ of the ``background'' FLRW space and with only two possible 
values of the parameter $\nu$. In the following we set $B_0=0$.

The Husain-Martinez-Nu\~nez geometry is asymptotically FLRW as  
$r\rightarrow +\infty$ and is exactly FLRW if $C =0$, in which case the 
constant $A_0$ can be  eliminated by rescaling the time $\eta$. The Ricci 
scalar is 
\be
{\cal R} = 8\pi \nabla^c\phi \nabla_c \phi = 
\frac{2\alpha^2 C^2 \left( 
1-\frac{2C}{r}\right)^{\alpha-2}}{
3 r^4  A_0 \eta } 
- \frac{ 3A_0^2}{ 
2 \left( A_0 \eta \right)^3  
\left( 1-\frac{2C}{r}\right)^{\alpha} } \,.
\ee
This Ricci scalar shows that there is a spacetime singularity at 
$r=2C$ (for both values of the parameter $\alpha$). The matter scalar 
field  $\phi$ 
also diverges there,  and there is a Big Bang singularity at $\eta =0$.

We have $ 2C <r<+\infty $  and the value  $r=2C$ corresponds to zero areal 
radius 
\be
R(\eta, r)= \sqrt{A_0 \eta} \, r \left( 
1-\frac{2C}{r}\right)^{\frac{1-\alpha}{2}} \,.
\ee
We can transform the time coordinate and use the comoving time $t$ of the 
``background'' FLRW space instead of the conformal time $\eta$ related to 
$t$ by  $ dt=ad\eta= \sqrt{A_0\eta}d\eta$ \cite{mybook}. Then we have    
\begin{eqnarray}
t &=&\int d\eta \, a(\eta)=\frac{2\sqrt{A_0}}{3} \, \eta^{3/2} \,,\\
&&\nonumber\\
\eta &=& \left( \frac{3}{2\sqrt{A_0}} \, t \right)^{2/3} \,,
\end{eqnarray}
and
\be
a(t) =\sqrt{A_0\eta}= a_0 \,  t^{1/3} \,.
\ee
The Husain-Martinez-Nu\~nez solution in comoving time reads
\be
ds^2 = - \left( 1-\frac{2C}{r}\right)^{\alpha} 
dt^2 +a^2(t) \left[ \frac{ dr^2}{\left( 
1-\frac{2C}{r}\right)^{\alpha} } 
+   \frac{r^2 d\Omega_{(2)}^2  }{ \left( 1-\frac{2C}{r}\right)^{\alpha -1} 
} \right] \,, 
\ee
\be
\phi( t, r ) = \pm \frac{1}{4\sqrt{\pi}} \, \ln \left[ 
D\left( 1-\frac{2C}{r}\right)^{\alpha/\sqrt{3}} 
a^{2\sqrt{3}}(t) \right] \,. 
\ee
The areal radius increases with $r$ for  $ r>2C$. In terms of the areal 
radius $R$, and using the notation 
\be
A(r) \equiv 1-\frac{2C}{r} \,, \;\;\;\;\;\; 
B(r) \equiv 1-\frac{(\alpha+1) C}{r}  \,,
\ee
we have 
\be
R(t, r)=a(t)r A^{\frac{1-\alpha}{2}}(r) \,.
\ee
A time-radius cross-term in the line element is eliminated by introducing  
a new time coordinate $T$ with differential  
\be
dT= \frac{1}{F} \left( dt+\beta dR \right) \,.
\ee
The choice  
\be
\beta(t,R)= \frac{ HRA^{\frac{3(1-\alpha)}{2}} }{
B^2(r)-H^2R^2 A^{2(1-\alpha)} } 
\ee
produces 
\begin{eqnarray}
ds^2 &=&- A^{\alpha}(r) \left[ 
1- \frac{H^2R^2 A^{2(1-\alpha)}(r)  }{B^2(r)} \right] 
F^2dt^2 +R^2 d\Omega_{(2)}^2 \nonumber\\
&&\nonumber\\
&\, & +\frac{H^2 R^2 A^{2-\alpha}(r)}{B^2(r)} \left[ 1+  
\frac{  A^{1-\alpha}(r)} { B^2(r) -H^2R^2  
A^{2(1-\alpha)}(r)}
\right] dR^2   \,.
\end{eqnarray}
The apparent horizons are located by the roots of the equation $g^{RR}=0$, 
which reads 
\be
\frac{1}{\eta} = \frac{2}{r^2} \Big[ r-(\alpha+1)C \Big] 
\left( 1-\frac{2C}{r} \right)^{\alpha -1}  \,.\label{equation}
\ee
As $r\rightarrow +\infty$, corresponding to $R\rightarrow 
+\infty$, Eq.~(\ref{equation}) reduces to $R\simeq H^{-1}$, the radius of 
the  cosmological apparent horizon in 
FLRW.  

Let $x\equiv C/r$, then the equation locating the  apparent horizons is 
\be
HR=\left[ 1-\frac{(\alpha+1)C}{r}  \right] \left( 
1-\frac{2C}{r} \right)^{\alpha-1} \,.
\ee
The left hand side of this equation is 
\be
HR=\frac{a_0}{3 \, t^{2/3}} \, \frac{2C}{x} \left( 1-2x 
\right)^{\frac{1-\alpha}{2}} \,,
\ee
while the right hand side is 
\be
\left[ 1-(\alpha+1)x\right](1-2x)^{\alpha-1} \,,
\ee
therefore the apparent horizons radii are expressed in parametric form by 
\begin{eqnarray}
t(x) &=& \left\{ \frac{2Ca_0}{3} \, \frac{ 
(1-2x)^{3(1-\alpha)}}{x\left[ 1-(\alpha+1)x\right]} 
\right\}^{3/2} \,, \label{forfig}\\
&&\nonumber\\
R(x) &=& a_0 \, t^{1/3}(x) \, \frac{2C}{x}\left( 1-2x 
\right)^{ \frac{1-\alpha}{2}} 
\end{eqnarray}
with parameter $x$. The numerical solution for the radii of these 
apparent horizons is shown in Fig.~\ref{fig:HMNfigure1}.

\begin{figure}
\centering
\includegraphics[width=11 cm]{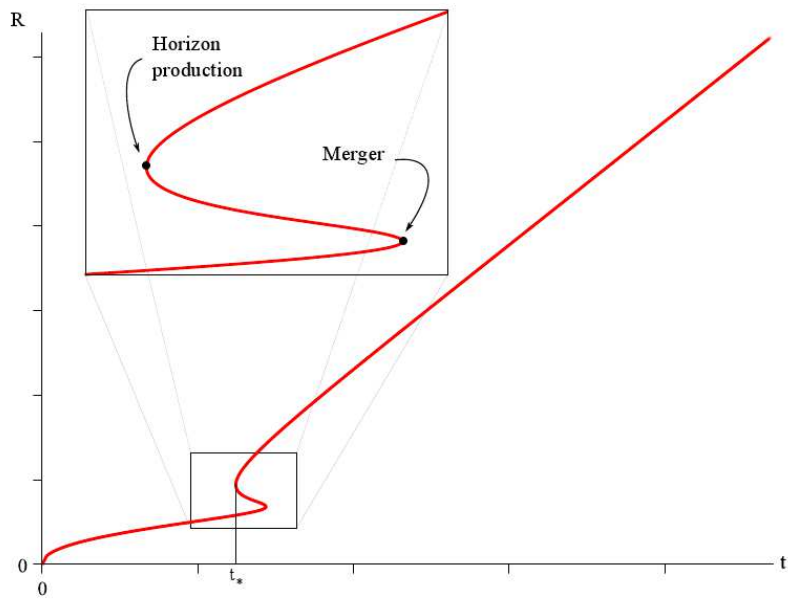}
\caption{The radii of the apparent horizons of the 
Husain-Martinez-Nu\~nez spacetime 
versus the FLRW comoving time for $\alpha=\sqrt{3}/2$ ($t$ and $R$ are 
in arbitrary units of length  and the 
parameter values  are such that 
$(Ca_0)^{3/2}=10^3$). 
\label{fig:HMNfigure1}}
\end{figure}

If $\alpha=\sqrt{3}/2$, between the Big Bang and 
a critical time $t_*$ there is 
only one expanding  apparent horizon, then two other 
apparent horizons are created at the critical comoving time $t_*$. One of 
them is  a 
cosmological  apparent horizon which expands forever, while the other is a 
black hole apparent horizon which contracts until it meets the first 
(expanding) black hole apparent horizon \cite{HMN}.  
When they meet, these two apparent horizons ``annihilate'' and a naked 
singularity appears at 
$R=0$ in a FLRW universe \cite{HMN, mybook}. This phenomenology of 
apparent horizons is dubbed ``S-curve'' behaviour from the shape of 
the curve representing the evolution of the apparent horizon radii  in 
Fig.~\ref{fig:HMNfigure1}. This ``S-curve'' phenomenology appears also in 
Lema\^itre-Tolman-Bondi spacetimes, in which a dust fluid curves the 
cosmological ``background'' \cite{BoothBritsGonzalez}. Multiple ``S''s are 
also possible ({\em e.g.}, five of them may appear in these 
Lema\^itre-Tolman-Bondi solutions). The scalar 
field is regular on the apparent horizons.

For the parameter value $\alpha=-\sqrt{3}/2$, there is only one 
cosmological  apparent horizon  and the 
universe contains a naked singularity at $R=0$.

The apparent horizons are {\em spacelike}: the normal vector to 
these surfaces always lies 
inside the light cone 
in an $\left( \eta, r \right)$ diagram \cite{HMN}, in agreement with a 
general result of Booth, Brits, and Gonzalez \cite{BoothBritsGonzalez} 
stating that a  trapping horizon created by a massless scalar field must 
be spacelike.

The singularity at $R=0$ is timelike for both values of the parameter   
$\alpha$. The central black hole in this spacetime is created with the 
universe in the Big Bang and not in a collapse process (although 
sometimes the literature states otherwise).

Clifton's 2006 solution of $f({\cal R} )= {\cal R}^n $ gravity 
\cite{Clifton06} (where ${\cal R}$ is the Ricci scalar) and some perfect 
fluid solutions of Brans-Dicke gravity found by Clifton, Mota, and Barrow 
\cite{CMB05} exhibit the same S-curve phenomenology of apparent horizons 
originally discovered in the Husain-Martinez-Nu\~nez spacetime \cite{VF09, 
 VitaglianoSotiriouLiberati12}. $f({\cal R})$ gravity is described 
by the fourth  order field equations 
\begin{equation}
R_{ab}-\frac{1}{2} \, g_{ab} {\cal R}= \frac{1}{f'( {\cal R})} \left[
\nabla_a \nabla_b f' -g_{ab} \Box f' +g_{ab} \, \frac{\left( f-{\cal 
R}f' \right)}{2} 
\right] \,,
\end{equation}
where $f'({\cal R}) \equiv df/d{\cal R}$. It can be shown that $f({\cal 
R})$ theories are equivalent to a subclass of Brans-Dicke theories with 
effective Brans-Dicke field $\phi=f'(R)$, coupling parameter  $\omega=0$, 
and the special (implicit) potential  
\be
V(\phi) = \phi {\cal R}(\phi)- f({\cal R}(\phi)) 
\ee
(see Refs.~\cite{review, DeFeliceTsujikawa, OdintsovNojiri} for reviews).

\subsection{The Fonarev and phantom Fonarev solutions}

Another inhomogeneous universe representing a central object embedded 
in a FLRW ``background'' is the Fonarev solution of GR \cite{Fonarev95}, 
which generalizes the Husain-Martinez-Nu\~nez spacetime (the latter has a 
free scalar field as the 
matter source) to the case in which the scalar field acquires the 
exponential  potential 
\be
V\left(\phi\right)=V_0 \, \mbox{e}^{-\lambda \phi} \,,
\ee
where $V_0$ and $ \lambda $ are positive constants. The 
coupled Einstein-Klein-Gordon equations are now
\begin{equation}
R_{ab}-\frac{1}{2} \, g_{ab} {\cal R}= 8\pi \left( \nabla_a\phi \nabla_b 
\phi -\frac{1}{2} \, g_{ab} \nabla^c \phi \nabla_c \phi -V(\phi) g_{ab} 
\right) \,,
\end{equation}
\begin{equation}
\Box \phi -\frac{dV}{d\phi} = 0 \,.
\end{equation}  
The Fonarev line element and scalar field are \cite{Fonarev95}    
\begin{eqnarray}
ds^2 & = & a^2 \left(\eta\right) \left[ -f^2\left(r\right) 
d\eta^2+\frac{dr^2}{f^2\left(r\right)}  
+S^2 \left(r\right) d\Omega^2_{(2)} \right] \,,\\
&&\nonumber \\
\phi \left( \eta, r \right) &=&
 \frac{1}{\sqrt{\lambda^2+2}}\ln
\left(1-\frac{2w}{r}\right) 
+\lambda\ln a
+ \frac{1}{\lambda} \ln \left[ 
\frac{V_0\left(\lambda^2-2\right)^2 }{2A_0^2 
\left(6-\lambda^2\right)} \right] \,, 
\end{eqnarray}
where
\begin{eqnarray}
f(r) &=&\left(1-\frac{2w}{r}\right)^{\alpha/2 } \,,\ 
\ \ \
\alpha =\frac{\lambda}{\sqrt{\lambda^2+2}} \,, \\
&&\nonumber\\
S(r)  & = & r 
\left(1-\frac{2w}{r}\right)^{\frac{1-\alpha}{2}  } \, ,\\
&&\nonumber\\
a(\eta) &=& A_0 |\eta|^{\frac{2}{\lambda^2-2}} \,,
\end{eqnarray}
and where  $w$ and $ A_0$ are constants, while $\eta $ is the conformal 
time of the FLRW substrate. Choosing $A_0=1$ and restricting to the 
parameter value  $w=0$, the Fonarev solution reduces to 
spatially flat FLRW. If  $a \equiv  1$ and $\alpha=1$, it reduces to the  
Schwarzschild geometry, while if $\lambda=\pm \sqrt{6}$ and $V_0=0$, it 
reproduces the Husain-Martinez-Nu\~nez spacetime.

A generalized Fonarev solution corresponding to  a phantom 
 scalar field solution of the Einstein equations was presented in 
\cite{GaoChenVFShen08} and is obtained from the 
Fonarev solution by means of the transformation
\begin{eqnarray} 
\phi\rightarrow i\phi \,, \ \ \ \ \lambda\rightarrow
-i\lambda \,.
\end{eqnarray}
The equation locating the apparent horizons is a quartic which  has only 
two real positive roots, corresponding to a cosmological  apparent horizon  
of radius $R_c$ and to a black hole apparent horizon of radius $R_b$ with 
the same qualitative 
behaviour of the apparent horizons of the (generalized) McVittie solution.

One can regard the Fonarev solution of GR as the Einstein frame version of 
a solution of Brans-Dicke gravity, which can then be mapped back to 
the Jordan frame\footnote{This is the technique used also by Clifton, 
Mota, and Barrow \cite{CMB05} to generate a conformal cousin of the 
Husain-Martinez-Nu\~nez spacetime which is a vacuum solution of 
Brans-Dicke theory studied in \cite{cousin}.} \cite{confonarev}. The 
conformal cousin 
of the Fonarev geometry is a 4-parameter family of solutions which are 
spherical, 
non-static, and asymptotically FLRW. The 
solutions of this family which have been studied explicitly 
contain only wormholes and naked singularities. Special cases of this 
4-parameter family provide solutions of vacuum $f( {\cal R})= {\cal 
R}^n$ gravity \cite{confonarev}.

\subsection{Other GR solutions}

Other better-known solutions of GR which may be interpreted as black 
holes embedded in cosmological substrata include  Swiss-cheese models 
\cite{EinsteinStraus1, EinsteinStraus2},  Lema\^itre-Tolman-Bondi black 
holes (with a 
dust-dominated FLRW ``background''), members of the large Barnes 
 family of solutions \cite{Barnes73}, and the Sultana-Dyer solution 
\cite{SultanaDyer}. Several 
other analytical  solutions of GR exist which describe central 
inhomogeneities in FLRW  ``backgrounds'' \cite{Fonarev95, Vaidya77,  
PatelTrivedi82, Roberts89, Burko97, Balbinot88, Cox03, Lindesay07, 
Lindesay13}. Many do not have reasonable matter sources and the 
energy density becomes negative in certain spacetime regions.  
Often these solutions are obtained by conformally transforming a  
stationary black hole solution or by performing 
a Kerr-Schild transformation of a stationary black hole metric $g_{ab}$, 
\be
g_{ab} \rightarrow \bar{g}_{ab}=g_{ab}+\lambda k_a k_b \,,
\ee
where $\lambda$ is a function of the spacetime position and the 
vector $k^a$ is null and geodesic with respect to both $g_{ab}$ and 
$\bar{g}_{ab} $ \cite{Krasinski, KrasinskiHellaby04, McClureDyer06, 
McClureDyer2, McClureetal07, McClureetal08}.

\subsection{Perfect fluid solutions of Brans-Dicke gravity}
\label{subsec:CMB}

Perfect fluid solutions of Brans-Dicke gravity describing dynamical 
cosmological black holes in certain regions of 
the parameter space were found by Clifton, Mota, and Barrow \cite{CMB05}.   
The Brans-Dicke field equations are
\begin{equation}
R_{ab}-\frac{1}{2} \, g_{ab} {\cal R}= \frac{8\pi }{\phi} \left[ 
T_{ab}+\frac{\omega}{\phi} \left( \nabla_a\phi \nabla_b\phi -\frac{1}{2} 
\, g_{ab} \nabla^c \phi \nabla_c\phi \right) +\nabla_a \nabla_b \phi 
-\frac{1}{2} \, g_{ab} \Box \phi \right] \,,
\end{equation}
\begin{equation}
\Box\phi=\frac{8\pi T}{\left( 2\omega+3 \right) \phi }  \,,
\end{equation}
where $\omega \neq -3/2$ is the Brans-Dicke coupling 
parameter and $T$ is the trace of the matter stress-energy tensor 
\begin{equation}
T_{ab}=\left( P+\rho \right) u_au_b +Pg_{ab} 
\end{equation} 
describing a perfect fluid with constant equation of 
state 
\be
P^{(m)}=\left( \gamma-1 \right) \rho^{(m)} \,,
\ee
while the Brans-Dicke 
field $\phi$ is free and massless. The line element of this 
family of solutions is spherical, inhomogeneous, and asymptotically FLRW:
\be
ds^2=-e^{\nu (\bar{r} )}dt^2+a^2(t) e^{\mu 
(\bar{r} )}(d\bar{r} ^2+\bar{r}^2d\Omega^2_{(2)}) \,,
\ee
where 
\begin{eqnarray} 
e^{\nu (\bar{r} )} & = &  
\left(\frac{1-\frac{m}{2\alpha \bar{r} }}{1+\frac{m}{2 \alpha \bar{r} 
}}
\right)^{2\alpha 
}\equiv A^{2\alpha} \,,\\
&&\nonumber\\
e^{\mu (\bar{r} )} & = & \left(1+\frac{m}{2\alpha \bar{r} }\right)^{4} 
A^{\frac{2}{\alpha}( \alpha-1)(\alpha +2)} \,, \\
&&\nonumber\\
\label{abeta}
a(t) & = & a_0\left(\frac{t}{t_0}\right)^{\frac{ 
2\omega_0(2-\gamma)+2}{3\omega_0\gamma(2-\gamma)+4}}\equiv 
a_{\ast}t^{\beta}  \,,\\
&&\nonumber\\
\alpha & = & \sqrt{ \frac{ 2( \omega_0+2 )}{2\omega_0 +3} } 
\,,
\end{eqnarray}
while the Brans-Dicke scalar field and the fluid energy density are 
\begin{eqnarray}
\phi(t, \bar{r} ) &= & 
\phi_0\left(\frac{t}{t_0}\right)^{\frac{2(4-3\gamma)}{ 
3\omega_0\gamma(2-\gamma)+4}}A^{-\frac{2}{\alpha }(\alpha^2-1)}\,,\\ 
&&\nonumber\\
\rho^{(m)}(t, \bar{r} ) & = & \rho_0^{(m)} \left( \frac{ 
a_0}{a(t)} \right)^{3\gamma} A^{-2\alpha} \,.
\end{eqnarray}
The areal radii of the apparent horizons of this class of solutions 
were studied in \cite{VitaglianoSotiriouLiberati12}, disclosing 
a rich variety of behaviours as  the three parameters 
vary. Some possibilities, corresponding to selected regions of parameter 
space, are shown in Figs.~\ref{fig:CMB1}-\ref{fig:CMB4} 
(we refer the reader to \cite{VitaglianoSotiriouLiberati12} for 
details).
 
\begin{figure}
\centering
\includegraphics[width=11 cm]{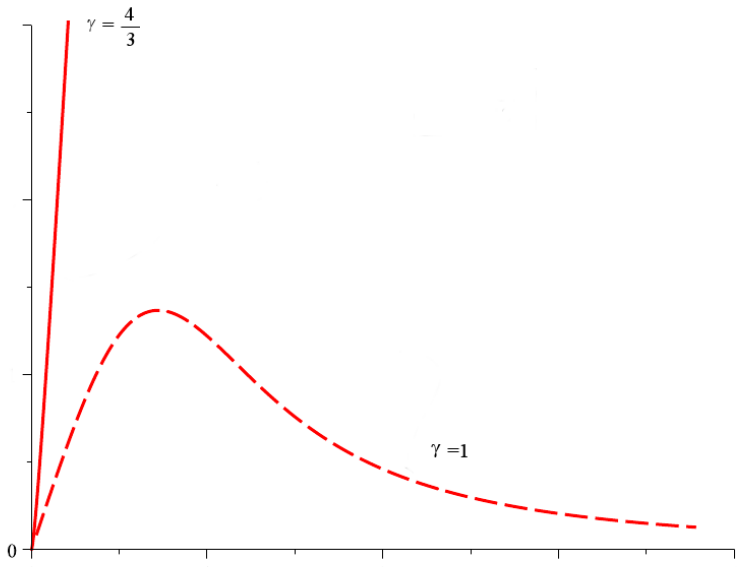}
\caption{The areal radii of the apparent horizons versus the FLRW comoving 
time 
(both in units of $(ma_0)^{1/(1-\beta)}$) for $\omega=-17/12$. The dashed 
curve corresponds to dust ($\gamma=1$) and the solid curve corresponds to 
both radiation ($\gamma=4/3$) and stiff matter ($\gamma=2$). For dust, a 
single apparent horizon expands to a maximum size and then shrinks.} 
\label{fig:CMB1}
\end{figure}

\begin{figure}
\centering
\includegraphics[width=11 cm]{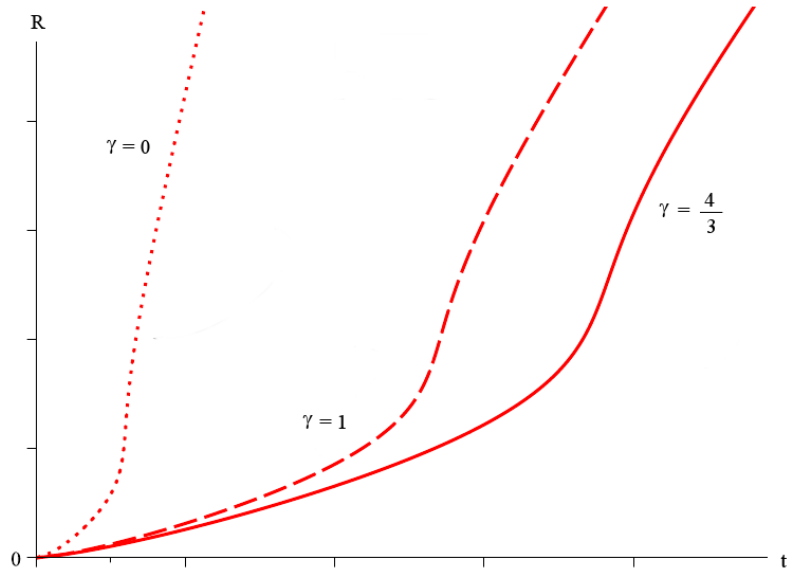}
\caption{Areal radii of the apparent horizons for Brans-Dicke 
coupling $\omega=-1/3$. 
The dotted curve corresponds to $\gamma=0$. In all three cases shown here, 
there are a single expanding horizon and a naked singularity.} 
\label{fig:CMB2}
\end{figure}

\begin{figure}
\centering
\includegraphics[width=11 cm]{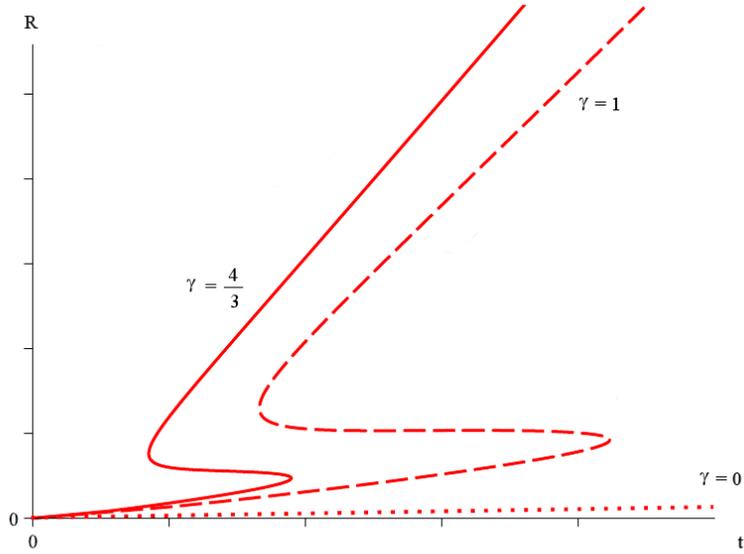}
\caption{The radii of the apparent horizons for Brans-Dicke 
coupling $\omega=1$. At early times 
there is a single horizon for all three 
values of $\gamma$. As time goes by, two more apparent horizons appear. 
Two of these horizons eventually merge and disappear, leaving a naked 
singularity in a FLRW universe, which has its own cosmological horizon. 
Fig.~\ref{fig:CMB4} zooms in on the third curve, which here appears 
flattened along the time axis.} 
\label{CMB3}
\end{figure}

\begin{figure}
\centering
\includegraphics[width=11 cm]{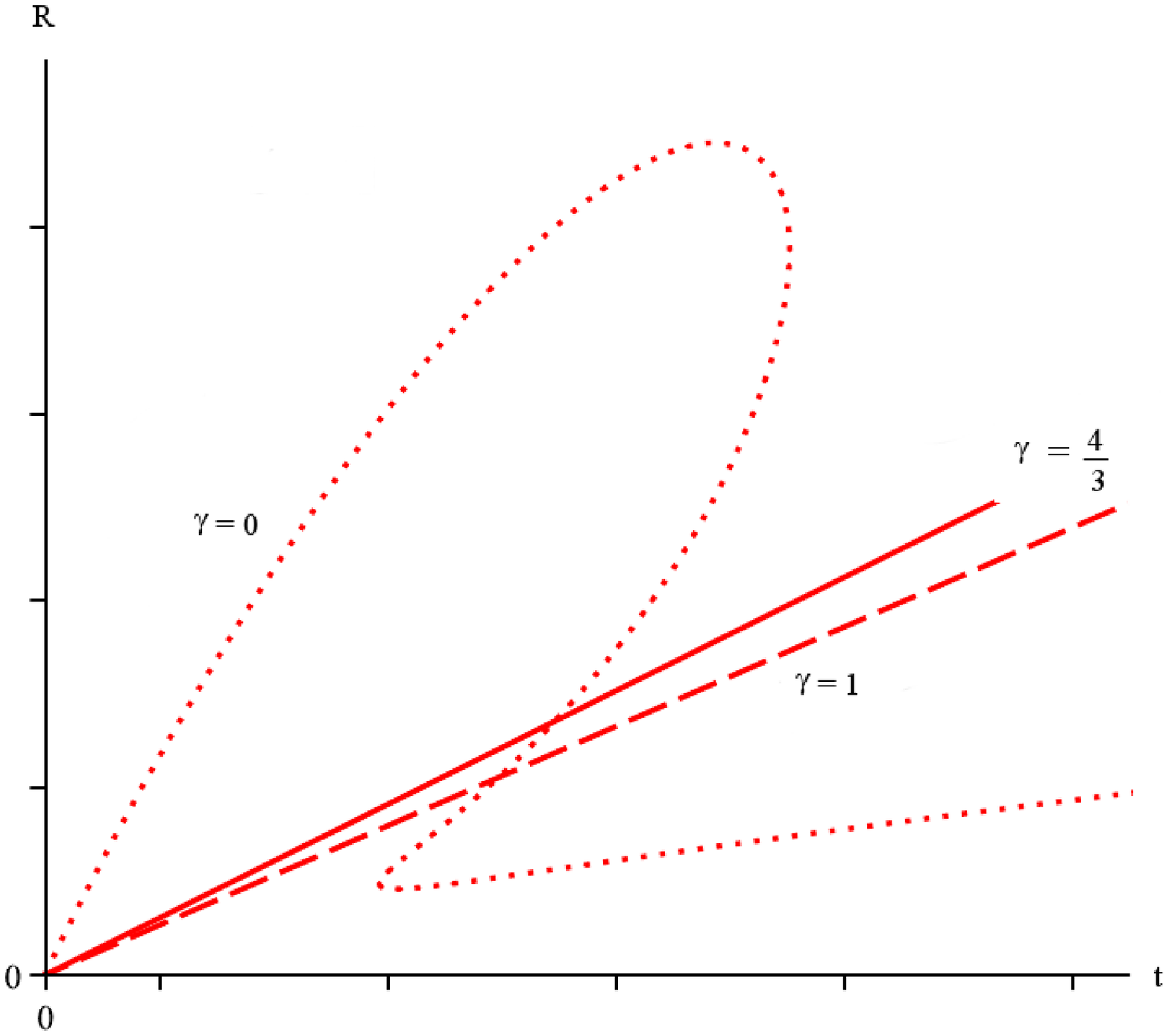}
\caption{A zoom of Fig.~\ref{CMB3} near the origin.}
\label{fig:CMB4}
\end{figure}

\subsection{Is there a relation between S-curve and C-curve?}

Thus far, we have encountered two main phenomenological classes describing 
the time behaviour of apparent horizons in spherical inhomogeneus 
spacetimes: the 
``C-curve'' behaviour exemplified by McVittie solutions and the 
``S-curve'' phenomenology originally discovered in the 
Husain-Martinez-Nu\~nez solution of GR. These two phenomenologies are 
encountered in GR and in other theories of gravity.  We know that other 
behaviours are in principle possible (cf. Sec.~\ref{subsec:CMB}) and that 
the catalog of analytic solutions with these physical properties is scarce 
in GR and even leaner in alternative theories of gravity. However, the 
tentative assumption that these two behaviours have some generality has 
some basis for the moment. One is then led to wonder whether the 
``S-curve'' and the ``C-curve'' behaviours are completely disconnected and 
mutually exclusive, or whether there may be some relation between them.  
As the Brans-Dicke coupling  
$\omega$ tends to infinity, the Clifton-Mota-Barrow solution of 
Brans-Dicke gravity discussed in 
the previous subsection  asymptotes to the comoving mass non-rotating 
Thakurta solution of GR \cite{mybook}. While the $\omega \rightarrow 
\infty$ limit of 
Brans-Dicke theory is expected to reproduce a corresponding GR solution 
\cite{Weinberg}, exceptions are known \cite{failure2, failure3, 
failure4, failure5, failure6, failure7} and often explained 
\cite{failure1, myBDlimit1, myBDlimit2}. In this case the non-rotating 
Thakurta 
solution is 
indeed the GR limit of the Clifton-Mota-Barrow solution, which was  
not realized nor expected {\em a priori}. But the most interesting thing 
is the behaviour of 
the apparent horizons of the Clifton-Mota-Barrow geometry in this limit: 
the S-curve describing the apparent horizons areal radii in the 
$\left( t, R \right)$ plane reduces to a C-curve as $\omega 
\rightarrow \infty$ because the lower 
bend of the S-curve is pushed to infinity \cite{VFPrain15}, as shown in 
Fig.~\ref{fig:limit}.

\begin{figure}
\centering
\includegraphics[width=11 cm]{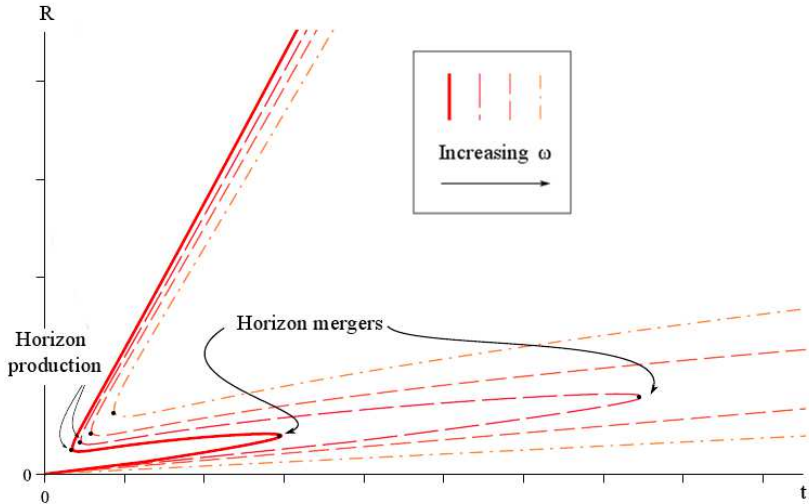}
\caption{The areal radii of the apparent horizons as the value of the 
Brans-Dicke coupling $\omega$ 
increases. The lower bend of the S-curve is pushed to infinity as $\omega 
\rightarrow \infty$.}
\label{fig:limit}
\end{figure}

Is the C-curve described by the apparent horizons radii of  a 
spherical, asymptotically FLRW spacetime always a limit of an 
S-curve? Does the relation between these two phenomenologies have any 
physical meaning or are we just looking at a coincidence? 
These questions are currently under investigation.

\section{Conclusions}
\label{sec:4}

To conclude this brief excursion on the subject of dynamical cosmological 
black holes and their relations with modified gravity, one should first 
realize that, while it is auspicable to find 
new inhomogeneous solutions amenable to this physical 
interpretation, it is usually 
harder to find whether apparent horizons exist, and to determine their 
location, nature, and dynamical behaviour, and some effort should be 
devoted to these goals. 

There are many open questions, some of which are more general than the  
subject of dynamical cosmological black holes, and are related to the 
nature of  horizons, which ultimately define the very concept of black 
hole. The main question seem to be whether apparent horizons are the 
``right'' surfaces to characterize black holes. It is a fact that apparent 
and trapping horizons are used in the numerical prediction of the 
waveforms of gravitational waves emitted by binary systems and used in 
their interferometric detection. The recent 
successes  
of the {\em LIGO} interferometers \cite{LIGO, LIGO2, LIGO4} in detecting 
the 
elusive gravitational waves from black hole mergers seem to 
answer the question affirmatively, but it is in principle possible that a 
different 
notion of horizon could be as successful as that of apparent/trapping 
horizon. In the meantime, the gravitational wave community  is not 
apologetic in using apparent and trapping horizons and in disregarding the 
teleological event horizon.

A different but related question, which has not been addressed here, is 
whether apparent horizons are the ``right'' surfaces to use in the study 
of the thermodynamics of dynamical black holes. There are strong 
claims that they are, but no convincing definitive proof has been 
provided. It is 
difficult to perform calculations of quantum field theory in curved space 
even on a prescribed time-dependent black hole background (that is, 
neglecting backreaction). The situation 
at the moment is summarized in the fact that the tunneling method provides 
a definite and general result for the Hawking temperature while other 
methods, at best, have produced results only for specific and very 
particular background metrics and a general computation cannot be 
completed as is done in the tunneling method. However, at present there 
is no guarantee that the result and 
procedure provided by the tunneling method are actually correct. Progress 
in this direction is very slow. At the very least, apparent horizons would 
require some adiabatic approximation if they are to make sense for 
dynamical black holes, otherwise one is probably discussing 
non-equilibrium thermodynamics, which is a tall order in the context of a 
problem which is proving to be already very difficult in equilibrium 
thermodynamics.

Assuming that apparent horizons are the ``correct'' notion of horizon, 
their biggest problem is no doubt their foliation-dependence, which 
amounts to the fact that the very existence of a black hole seems to 
depend on the observer. This problem is exemplified dramatically by the 
realization that there exist slicings of the Schwarzschild spacetime with 
no apparent horizons \cite{WaldIyer91, SchnetterKrishnan06}. The fact that 
these slicings are rather contrived is not of much consolation. From the 
practical point of view, the 
problem is mitigated by the fact that, in spherical symmetry, the apparent 
horizons coincide in all spherically symmetric foliations 
\cite{VFEllisFirouzjaeeHelouMusco17}, but this fact does not really solve 
the problem and, moreover, no analogous result has been proved beyond 
spherical symmetry.

In our examples of apparent horizon phenomenology, we have restricted 
ourselves to spherical symmetry. Evolving apparent horizons exhibit a 
rather rich phenomenology and dynamics, but there seem to be two main 
phenomenological classes for the behaviour of the areal radii $R_{AH}(t)$ 
of apparent horizons versus the comoving time of the FLRW substratum in 
which the spherical inhomogeneities are embedded. These are the 
``C-curve'' first encountered in the McVittie spacetime of GR 
\cite{McVittie33} and the ``S-curve'' first discovered in the 
Husain-Martinez-Nu\~nez solution of the same theory \cite{HMN} and then 
rediscovered in $f( {\cal R})$ and Brans-Dicke gravity \cite{VF09, 
VitaglianoSotiriouLiberati12}. There are hints 
that some relation may exist between these two classes, with the C-curve 
being some limit of the S-curve, but the situation is not clear at 
present.

A final question is provided by the long-standing issue of cosmic 
expansion versus local dynamics which motivated the introduction of the 
McVittie metric in 1933 and of the Einstein-Straus Swiss-cheese model a 
decade later. This problem of principle, which at some point was brushed 
off as being irrelevant, keeps being discussed in the literature (see 
\cite{CarreraGiulini} for a 2010 review). We have seen already in the few 
examples discussed here that sometimes black hole apparent horizons 
expand 
(in rare cases they are even comoving with the cosmic substratum), whereas 
some other times they resist the cosmic expansion or even contract. No 
general rule appears here and the apparent horizons of dynamical 
cosmological black holes provide no definite answer to the puzzle. All 
these theoretical questions of classical gravity are worth understanding 
and investigating in the future. The recent detection of 
gravitational waves and the future development of gravitational wave 
astronomy seem to reformulate the issue of apparent horizons in more 
practical terms and to make its understanding a more pressing issue than 
it was before.

\section*{Acknowledgments} 

The author thanks Angus Prain for producing an earlier version of most 
figures, a referee for helpful comments, and Bishop's University for 
support.

\end{document}